\documentstyle[editedvolume,namedreferences]{crckapb}
\begin{opening}
\title{The Square Kilometer Array: new challenges for cosmology}
\author{Rien van de Weygaert and Tjeerd S. van Albada}
\institute{Kapteyn Astronomical Institute, Groningen, the Netherlands}
\end{opening}
\runningtitle{The SKA: new challenges for cosmology}
\begin{document}
\vskip -1cm
\centerline{\it To appear in:}
\centerline{``The Westerbork Observatory; Continuing Adventure in
Radio Astronomy''}
\centerline{\it Eds. E. Raimond \& R. Genee (Kluwer, Dordrecht)}

\section{Introduction}

In the past three decades our view and understanding of the structure 
and evolution of the universe has grown in an almost revolutionary 
fashion. The discovery of the rich variety of patterns displayed by
the galaxy distribution, the discovery of streaming motions in the
local universe, the detection of tiny variations in the temperature of 
the microwave background radiation and the use of powerful
computational resources for simulating the formation and evolution of 
structure are some of the most conspicuous developments that
transformed the field of cosmology into one of the most active 
branches of astrophysics\footnote{for up-to-date discussions and overviews 
on cosmology see e.g. Kolb \& Turner 1990; Peebles 1993; and Jones
1994; excellent texts that focus specifically on the issue of large 
scale structure and structure formation are Padmanabhan 1993 and 
Coles \& Lucchin 1995; Peebles 1980 is the standard reference.}. 
It is therefore plausible that the Holy Grail
of cosmology -- the answer to the question of how the wealth and
variety of structure in our cosmos rose out of an almost featureless, 
extremely hot and dense early universe -- may soon come within reach. 

Twenty-five years ago, the prospect of measuring the structure of the
universe 
was one of the main motivations for building the Westerbork
Synthesis Radio Telescope. In particular, it was hoped that
by counting radiosources as a function of their brightness it would be
possible to determine the cosmological parameter $q_{0}$, which
measures the deceleration of the expansion of the universe. That this
expectation has not been fulfilled was due to several factors. 
The main reason was that the surveyed volume did not quite reach 
out to the large redshifts necessary for discriminating between different 
values of $q_{0}$. Moreover, the 
interpretation of the data got insuperably complicated by the rich 
variety of radiosources and their evolution. In combination with 
fascinating discoveries such as radiosources with jets this lead to 
a shift of research interests. In the developments in the 
subsequent golden era of cosmology radio observations therefore hardly staged 
more than a background role, although quite often a significant one.  

With the upgrades of the Westerbork Telescope (WSRT) and the VLA, and in 
particular when the ambitious plans for a Square Kilometer Array
become reality, it is 
quite likely that radio astronomy will play a keyrole in the cosmological
developments of the coming decades. This contribution
\footnote{We should note here that this contribution is not intended 
to be an extensive review, and we therefore apologize in advance for failing 
to cite all relevant and important contributions to the items discussed.}
is meant to give 
an impression of some of the main cosmological issues in which the SKA
could play a large role and to whose solution the SKA could contribute
significantly. Firstly, in section 2 we will give a general inventory
of possible cosmological applications, some of which we will discuss
in more detail. Of great interest is the use of the 
SKA for carrying out redshift surveys on the basis of the 21 cm line. 
In section 3 we will therefore focus in particular on this aspect 
of galaxy redshift surveys, and describe in more detail what the
present knowledge of the morphology and geometry of the galaxy
distribution is, and why the SKA would be such a magnificent instrument 
for mapping the galaxy distribution to high redshifts, and therefore 
to provide important contributions towards understanding the formation
of structure in the universe. In the two subsequent sections we  
discuss the technical issues involved with the use of the SKA for 
the described cosmological applications. In particular we will address
the number of gas-rich galaxies that we can hope to detect and the 
observing time required to detect a given amount of hydrogen gas at 
some redshift $z$. On the basis of these considerations we then
conclude with specifying the requirements that would make the 
Square Kilometer Array into an ideal instrument for cosmology.

\section{The Square Kilometer Array: Inventory of its cosmological 
potential}

It have been mainly optical telescopes that were instrumental in 
mapping the large-scale structure of the galaxy distribution. By
virtue of the Hubble expansion of our universe we can determine the 
distance of each galaxy 
simply by measuring the redshifts of the spectral lines in their 
electromagnetic 
spectrum. For the large majority of galaxies this has been done for lines 
in the optical region of their spectra. As yet the potential of 
determining the redshifts of galaxies with radiotelescopes
-- from the 21 cm (1420 MHz) line emitted by neutral atomic hydrogen 
(HI) -- has not been explored and applied over a large volume of the 
universe. This is mainly due to the limited frequency coverage and 
insufficient sensitivity of present-day radio emission receivers.

In the context of galaxy redshift surveys it is important to know 
that in the local universe essentially all neutral atomic hydrogen is 
associated with galaxies. There is no evidence for the existence of 
a major intergalactic component of HI at low redshifts, although 
this situation may have been considerably different at earlier epochs. 
At higher redshifts a large fraction of the hydrogen gas may actually 
have resided in intergalactic cloud complexes whose content accreted 
upon (proto)galaxies. On the other hand, because their stars have
formed out of their supply of hydrogen gas, the hydrogen content of
galaxies must have been larger in the past than it is now. Another 
significant point of 21-cm line redshift
surveys is that they concentrate specifically on spiral and irregular 
galaxies. These late-type galaxies contain substantial amounts of HI
gas and are therefore easily detectable by radio telescopes. On the
other hand, radio telescopes can not be used to detect and map the 
distribution of elliptical and SO galaxies, as they do not contain
enough atomic hydrogen. These early-type galaxies form only a small
minority of the total galaxy population, residing preferentially in 
high-density regions such as clusters of galaxies. The spiral and 
irregular galaxies, representing the major share of the galaxies 
in the universe, constitute a considerably less biased tracer of the
distribution of (luminous) matter, and they can be found in high as
well as low density regions. Moreover, an important general property
of these late-type galaxies is that their relative atomic hydrogen
content increases with decreasing optical luminosity (see e.g. Briggs
1990). As a result, `dwarf' galaxies are far easier to detect with
radio telescopes than with optical telescopes. Because the number 
of dwarf galaxies far exceeds the number of giant galaxies, the 
mapping of their distribution is essential for obtaining a complete
picture of the large scale distribution of galaxies in space.  
Radio redshift surveys therefore not only provide a useful and efficient
additional way of measuring galaxy redshifts, for the purpose of 
`cosmography' they also form a necessary complement to the optical
redshift surveys.  

So far most 21-cm spectroscopy work has been carried out with single-dish 
telescopes. In 
particular the Arecibo antenna has been instrumental in obtaining 
21-cm line redshifts, of which by now there have been measured some 
12,000 (see Giovanelli \& Haynes 1991). Single-dish radio telescopes, 
however, cannot compete against radio interferometric instruments in 
spatial resolution and efficiency. In this context it is worthwhile to
emphasize the interesting advantage that redshift surveys with radio 
aperture synthesis instruments will have over their optical 
counterparts. At optical wavelengths, redshifts are basically
determined `galaxy by galaxy', after they have been identified on the
sky. The same is true for redshift determinations with single-dish 
radio telescopes. With multi-dish radio interferometer telescopes on
the other hand, a predefined region of the sky gets `mapped' in small 
frequency bands that together cover a wide frequency range. This
spectral `mapping' capability of such radio telescopes is the key to 
a highly efficient method of determining redshifts - making it
possible to obtain the 21-cm line profiles and redshifts for huge
numbers of galaxies in the same observation. The coming upgrade of the 
receivers of the Westerbork array (WSRT) in 1997 will therefore 
undoubtedly lead to a significant improvement of the radio redshift 
survey capacity and enable comprehensive radioastronomical studies of 
the structure and kinematics of the local universe. 

An even more dramatic addition to our understanding of the morphology
and evolution of structure in the universe can be expected 
from the construction of a Square Kilometer Array (SKA). 
As this instrument will surpass any existing radio telescope 
by two orders of magnitude in sensitivity it would be difficult,  
taking into account the historical background, to try to predict  
how radio cosmology based on such an advanced instrument would 
develop in the next decades. It would even be difficult to aim 
at a complete and detailed scientific justification for such an 
instrument. Nonetheless, given this caveat, it is of course a challenging 
exercise to contemplate what an instrument like the SKA might 
contribute to the study of the large-scale structure of the universe. 

In this contribution we will outline in more detail the cosmological 
background and issues for which the SKA promises to be such an 
extremely powerful and important instrument. Indeed, we will argue 
that observations with the SKA will often be essential for 
significant advances in studying the global properties of the universe
as well as its `large scale structure' on megaparsec to gigaparsec
scales. In this context we will in particular focus on the possibility
of studying the neutral hydrogen gas in and around (and in between) 
galaxies out to large redshifts. Gas masses characteristic 
for galaxies similar to our own will probably be detectable out 
to redshifts of around $z \sim 3$. This will most likely 
be a fairly sharply defined redshift cut-off. It would demand 
too much effort to attempt to detect neutral hydrogen beyond 
this redshift threshold, because the expansion of the universe 
leads to an even stronger decrease of brightness with distance 
than the usual geometric fall-off. But inside that enormous volume 
hundreds of millions of galaxies can in principle be detected, which 
will allow a detailed cartography of the large and smaller scale
features in the distribution of luminous matter as well as an accurate
diagnosis of the value of the global cosmological
parameters. Throughout we will make the implicit assumption that the 
evolution of the gas clouds will be not so substantial that they would
severely degrade the cartographical capacity of the SKA. In
particular, we will assume that ionisation of hydrogen by quasars at
high redshift will not lead to a drastic reduction of the amount of
neutral hydrogen. In this respect it is comforting that absorption line 
studies of quasar spectra indicate that there are large numbers of HI
clouds at high redshifts. 

The fact that an instrument like the SKA would enable us to chart out 
the distribution of essentially {\it all} late-type galaxies over the 
largest part of the visible universe implies that 
it would also allow us to study the {\it evolution} in 
the structure of the universe, over a time interval corresponding to 
80 or 90 per cent of its age. Mapping out galaxy positions to
presently almost inconceivable redshifts of $z \sim 2.5$--$3.0$ 
means that we will have a look at the structure of the universe 
when it was only some 2 billion years old, compared to its present age
of about 12 billion years. Besides redshift surveys, also other 
cosmologically relevant observations will greatly benefit from the 
SKA. One such key observation is the mapping of the velocity field 
in the local universe, of crucial importance in obtaining a better 
understanding of the dynamics underlying the structure formation 
process. Reasonably accurate estimates of the peculiar velocities of spiral 
galaxies with respect to the expanding universe can be obtained 
via the Tully-Fisher relation, which links a spiral galaxy's rotation 
velocity to its luminosity. With the help of the SKA it will be possible to 
considerably extend the region in which one can measure these 
peculiar velocities. This would for example open up the exciting 
possibility of studying in detail the dynamics of structural features 
like `filaments', `great walls' and `voids'. 

Another issue on which the SKA will shed new 
light is that of the formation and evolution of galaxies. The 21-cm
line profile not only yields information on the redshift of these
galaxies, but also on their structure and kinematics. By virtue of the imaging 
capabilities of the SKA, coupling high sensitivity to a good
resolution, it will be possible to investigate the
kinematics and morphology of gas-rich galaxies out to high
redshifts, $z \sim 1$ and probably even higher. At a redshift as 
high as 4 the first galaxies may well have formed already, that is, 
individual gas clouds may have converted part of their hydrogen gas
into stars. The SKA will make it feasible to study the changes in the 
atomic hydrogen content of the galaxies due to the formation of stars and 
other astrophysical processes, like the formation and evolution of
galaxy discs, over a time period of nearly 10 billion years, i.e. over
almost the entire history of the universe (except of course for the
crucial first few billion years). The SKA will therefore not only
allow a study of the evolution of the large-scale distribution of
galaxies but also a (statistical) study of the evolution of galaxies 
themselves. 

Observations with the SKA may also lead to important contributions
towards an accurate and unbiased determination of the global
cosmological parameters. Better 
estimates of the expansion rate of the universe -- the Hubble
parameter\footnote{in this contribution we write the 
Hubble constant to be $100h\,\hbox{km/s/Mpc}$} $H_0$ and $q_0$ -- will be
possible through the application of 
the Tully-Fisher relation to galaxies at great distances. Moreover, if
the SKA would be provided with receivers in the Gigahertz spectral
region it would be an ideal instrument for imaging the
Sunyaev-Zel'dovich effect --- the distortion of the microwave
background spectrum due to the Compton scattering of microwave 
backgrounds photons by hot gas in galaxy clusters. Estimates of the 
value of the Hubble parameter, as well as that of the cosmic deceleration 
parameter $q_0$, follow out of the combination of these observations 
with X-ray measurements of the same clusters. They have the important 
advantage that they are not plagued by contaminating 
effects of locally induced peculiar velocities. 
Another interesting possibility of determining
the Hubble parameter via radio astronomical 
work involves the search for gravitationally lensed radio ring images of 
background extended radio sources in a suitably large and deep radio 
survey. These rings are a consequence of the general relativistic 
effect of gravitational focussing of light by a mass concentration 
like a cluster of galaxies. As the configuration that leads to the 
formation of radio rings is so highly constrained, it would be possible to
infer a value of $H_0$ without too many underlying assumptions. 
By applying classical cosmological tests like source counts, more so than 
via the Sunyaev-Zel'dovich effect, the SKA is expected to be of great 
value in measuring the rate at which the expansion of the universe
slows down. It is quite feasible that radio source counts with the SKA -- in
particular involving galaxy samples of which also the redshift can 
be determined and of which the evolutionary properties are known 
through the imaging capabilities of the telescope -- will go down to 
such faint flux levels (flux = observed brightness) that it finally will
become possible to determine $q_0$, and therefore the matter content
of the universe, quantified by $\Omega_0$. It may even be possible to
get a reliable estimate (or upper limit) to the cosmological constant 
$\Lambda$. 

\section{Cosmological Cartography: the large scale structure and 
morphology of the local universe}

When averaged over enormous scales of hundreds of megaparsecs the 
universe appears to be rather uniform. On smaller scales on the 
other hand, the salient features displayed by the distribution of 
galaxies testifies of another story (e.g. fig. 1). We see a baffling
variety of structures over a wide range of scales within the limited
realms of the local universe wherein structures have been mapped in
some detail. Galaxies tend to clump in a hierarchy of ever larger 
agglomerations, ranging from small groups of a few dozen galaxies up
to rich clusters of galaxies containing several hundreds to thousands
of them, which in their turn often group in superclusters that
sometimes house several tens of clusters and many more small groups. 

Whether the wealth of structure observed in the distribution of
galaxies also reflects the structure of the underlying matter 
distribution is a not yet completely settled issue. But if we 
wish to use the galaxy distribution to infer information on the 
large scale matter distribution and the structure formation process 
we must make that assumption. Fortunately, observations have 
indicated it to be plausible.

\subsection{Galaxy redshift surveys}

A first requisite for studying the galaxy distribution is to map their
positions over a large part of the universe. A lot of information on
the clustering of galaxies can already be obtained from sufficiently 
well-defined surveys of the galaxy distribution on the sky. They 
have the advantage that it is relatively straightforward to map 
the positions of a huge number of galaxies. In particular noteworthy 
in a historical context is the famous Lick map constructed by 
Shane \& Wirtanen (1967), covering nearly the whole sky and containing around
a million galaxies with an apparent magnitude $m<19$. Even better
and more objectively defined surveys have recently been obtained by 
using automatic plate-measuring machines. In particular the 
Edinburgh-Durham Southern Galaxy Catalogue (Heydon-Dumbleton, Collins
\& MacGillivray 1989) and the APM catalogue (Maddox et al. 1990a;
Maddox, Efstathiou \& Sutherland 1990b) have been 
very useful for cosmological research. For example, the APM catalogue 
contains about two million galaxies with $m < 20.5$ located in a
region of 4300 square degrees on the southern sky, probing to an
effective depth of around $600h^{-1}\,\hbox{Mpc}$. 

However, the sky catalogues do not contain distance information, and 
are therefore very limited when one addresses the issue of the
patterns in the galaxy distribution. Fortunately, the expansion 
of the universe provides us with a good and easy to measure estimate
of a galaxy's distance, the redshift $z$ of the electromagnetic
spectrum of the light emitted by the galaxy due to its recession 
velocity $V_0$. In the first half of this century Hubble discovered
that the velocity $V_0$ with which an object speeds away from us is directly
proportional to its distance $r$, $V_0 = H_0 r$. The precise value of 
the expansion rate -- the Hubble parameter $H_0$ -- is not yet known, 
although significant progress has recently been made on the basis of 
observations with the Hubble Space Telescope (on the basis of 
observations of Cepheids in M100 and M96 $H_0$ has been estimated 
to be in the order of $70-80$ km/s/Mpc, Freedman et al. 1994). In
addition, one should realize that the redshift of a galaxy not
only relates to its cosmic expansion velocity but that it also
includes its peculiar velocity with respect to the expanding
background universe (see next section). Although the latter usually 
represents only a minor contribution, redshifts should be taken 
merely as first estimates of distance. Because it is relatively 
easy to determine redshifts, with the arrival of very efficient
spectrographs and multiple object spectrographs it became feasible to 
chart large areas of the local universe. For statistical studies 
it is important to use well defined galaxy samples.
Usually the survey objects are identified from sky surveys 
on the basis of some selection criterion, often the apparent 
brightness determined from optical sky survey plates, but also for 
example the infrared or X-ray flux measured with a satellite.
To get an idea of the patterns in the large scale galaxy distribution,
and their geometry, the redshift surveys should also subtend a 
sufficiently wide solid angle on the sky. Such systematic wide angle 
redshift surveys have become our primary source of information 
for these purposes. Although early redshift compilations like 
the one by Humason, Mayall \& Sandage (1956) and the Revised 
Shapley-Ames Catalogue of nearby Galaxies (Sandage \& Tammann, 1981) 
already contained a wealth of information on structures in the nearby
universe, the compilation of comprehensive wide angle redshift surveys
really started in the 1980s with the famous Harvard-Smithsonian Center
for Astrophysics (CfA) survey by Geller, Huchra, and collaborators
(see e.g. Geller \& Huchra 1989). In the meantime numerous other 
surveys have become available, and some very ambitious optical surveys
are in an advanced stage of preparation, pushing back the boundary 
of the mapped universe. Figure 1 gives an
impression of the large-scale structure as we know it today, showing 
the galaxy positions in the CfA2 redshift survey and its southern 
counterpart, the SSRS2 survey (da Costa et al. 1994). Plotted are 
the right ascension versus the recession velocity of each of the 
15,000 galaxies in the sample, having an outer boundary of the
volume surveyed lying at a recession velocity of about 15000 km/sec. 
On the basis 
of these and other available surveys we can make a tour along the 
various structures to be found in our local niche of the universe. 

\begin{figure}
\vspace{15.0truecm}
\caption{The galaxy redshift distribution (da Costa 1993), right
ascension versus recession 
velocity, in opposite $18^{\circ}$ thick slices (in the indicated
declination range) in the northern CfA2 survey (see e.g. Geller \&
Huchra 1989) and the southern SSRS2 survey (da Costa et al. 1994). 
All galaxies with a magnitude brighter than $m=15.5$ are included. The
northern slice includes the Coma cluster and the Great Wall. In the
southern slice several wall-like structures can be seen forming a 
cellular-like structure (also see da Costa 1993). Figure kindly 
provided by L. da Costa.}
\end{figure}

\subsection{Inventory of nearby cosmic structures}

Our immediate neighbourhood provides a reasonably representative 
sample of the universe, with a distribution that is evidently 
far from uniform. Our own Galaxy has two nearby companions --  the
Large and Small Magellanic Clouds -- and together with the Andromeda 
galaxy M31 it dominates the so-called {\it Local Group} of galaxies, 
which consists of some 30 members and is a few Mpc in extent 
(Hodge, 1994). Further away, at a distance of approximately 17 Mpc, 
we encounter a prominent cluster of galaxies, the {\it Virgo cluster}. Its   
gravitational pull induces an infall velocity of a few 
hundred km/s of the Local Group towards the Virgo cluster. 
Although its proximity facilitates a detailed study of its 
structure, it is rather difficult to get an overall view of the
cluster because of its large angular extent on the sky. Some 1300 
members, and 500 additional possible members have been 
identified. A detailed analysis (Binggeli, Tammann \& Sandage 1987)
shows it to 
be a rather irregular structure with two pronounced concentrations, 
of which the major one contains the giant ellipticals M87 and M49. 
Several spatial and kinematic characteristics of the cluster suggest 
it to be a rather young structure, among others the fact that 
there is quite some substructure in the vicinity of the cluster 
core, which therefore must still be in the process of formation. A 
property that
the Virgo cluster has in common with many other clusters is the strong 
spatial segregation, with the early type galaxies being considerably 
more concentrated to the cluster centres than the spirals and 
irregulars. Within the scheme of galaxy clustering and structure 
formation the dense and rich clusters of galaxies stand out as the structures 
that represent the fringe of objects that can still be considered 
individually distinguishable entities, in the sense of being the most 
massive fully collapsed objects in the universe. Wandering further 
out from the Virgo cluster, we find several substantially more
impressive clusters within $100h^{-1}$ Mpc of the Local Group. One of
the most massive clusters we know of, the {\it Coma cluster} at a distance
of around $68.5h^{-1}$ Mpc, is located within this region (see
fig. 2). It is one
of the most regular, richest and best-observed clusters, containing many 
thousands of galaxies. Its regular spherically symmetric appearance 
and relative proximity made the Coma cluster into the canonical 
example of a rich and relaxed cluster, against which the
characteristics of all other clusters were compared. However, recent 
work, especially concerning the X-ray emission associated with 
the hot intracluster gas that has been trapped in its potential 
well, has indicated that the cluster displays significant  
substructure (White, Briel \& Henry 1993). The presence of such substructure
is a strong indication 
that clusters are relatively young and dynamically active 
objects. A recent close examination of the galaxy distribution and 
kinematics of the Coma cluster confirmed this, showing that the 
cluster consists of a massive central cluster and 
a clearly distinguishable smaller subcluster that is just falling 
onto Coma, while the core itself consists of at least two subclumps 
that are already substantially disrupted and in the process 
of merging (Colless \& Dunn 1995).

\begin{figure}
\vspace{11.5truecm}
\caption{Optical image of the central region of the Coma cluster. The 
cluster contains thousands of galaxies. 
The picture is obtained from the Digitized Sky Survey, based on photographic
data of the National Geographic Society -- Palomar Observatory Sky 
Survey (NGS-POSS), and was kindly provided by H. B\"ohringer.}
\end{figure}

Also, the Coma cluster offers a detailed view of the relation of
clusters to their spatial surroundings. Coma is not an isolated
object. On the contrary, going outward from the cluster core we see a 
gradual transition of an increasingly flattened galaxy distribution 
into a strongly elongated filament, in particular when looking 
at the distribution of early-type galaxies (Doi et al. 1995). This 
filament extends from the NE to the WSW and embeds the Coma cluster
into the surrounding large-scale structure. Since the publication of 
the results of the CfA survey (e.g. Geller \& Huchra 1989)
we know that 
the Coma cluster is in fact the strongest density concentration in 
the ``Great Wall'', a gigantic and coherent plane-shaped assembly 
of galaxies, whose dimensions are estimated to be of the order of 
$60h^{-1} \times 170h^{-1} \times 5h^{-1}\,\hbox{Mpc}$. This 
structure can be clearly discerned in figure 1 as the dark and thick band of 
galaxies in the northern hemisphere, running all the way from the left
to the right of the survey slice at a recession velocity of around 
$7,000\,\hbox{km/s}$. The Coma cluster itself is visible as the 
``Finger of God'' in the centre, the elongated stripe pointing towards
us in the redshift map induced by the dispersion of peculiar
velocities of its galaxies. More detailed 
studies reveal that besides the Coma cluster itself several other 
density enhancements can be recognized within this wall 
of galaxies. An example is the A1367 cluster, lying along the
direction of the the filament emanating from the Coma cluster. 

In fact, the relation and proximity to other clusters contains 
a lot of information on the presence and extent of even larger 
structures in which the clusters themselves are embedded. Numerous 
studies have shown that the distribution of rich clusters forms a
useful and to some extent complementary tracer of the large scale
matter distribution in the universe, certainly for scales 
exceeding twenty or more megaparsec (see Bahcall 1988). In particular 
the well-known
catalogue of the most prominent clusters compiled by Abell (1958), though 
probably beset by some not completely understood artefacts, has been 
of great use in obtaining an idea of clustering on such scales. 
From Abell's catalogue as well as from some more recent compilations 
of clusters of galaxies we have learned that they have the tendency 
to group even closer together, relatively speaking, than galaxies 
themselves. Together with groups
of clusters and a host of more loosely bound galaxies they appear to 
congregate in superclusters. These {\it superclusters} are huge,
loosely-bound, non-virialized structures containing several to many 
rich clusters grouped in a usually highly flattened or elongated
configuration. In comparison to galaxy clusters they represent far
smaller overdensities, usually in the order of only a few times the 
average density of the universe, so that they did not yet have the time 
to collapse. Their structure and dynamics are therefore still closely 
related to the initial density fluctuations that gave rise to them, 
making them interesting probes, `fossils', of the structure formation 
process. It actually took some time before the reality of these
structures got accepted, and it was not until the review by Oort
(1983) that they indeed got generally recognized as such. 

\begin{figure}
\vspace{11.0truecm}
\caption{The Perseus-Pisces supercluster chain of galaxies. Separate 
two-dimensional views of the galaxy distribution in the northern 
region of the Pisces-Perseus region. The upper panel shows the sky 
distribution of all galaxies in the overall northern survey sample 
of Wegner, Haynes \& Giovanelli (1993). The region believed to contain
the Pisces-Perseus main ridge is outlined. The lower panel shows the
two dimensional redshift distribution (right ascension-recession 
velocity $V_0$) for galaxies in the ridge region highlighted in the 
upper panel. From Giovanelli \& Haynes 1996, kindly provided by 
M. Haynes.} 
\end{figure}

Both our Local Group and the Virgo cluster are members of such a 
structure, the Local Supercluster, a huge flattened concentration of
about fifty groups of galaxies in which the Virgo cluster is the
dominant and central agglomeration. The Local Supercluster is in fact
a modest specimen of its class, dominated by only one rich cluster. 
A far more prominent example of a supercluster, and in some sense more
characteristic if one wishes to obtain an idea of their morphological 
variety, is the Perseus-Pisces supercluster (see fig. 3). Due
to its relatively nearby distance of around $55h^{-1}\,\hbox{Mpc}$,
its characteristic and salient filamentary geometry, and the main 
ridge's favorable orientation perpendicular to the line of sight, it 
has become one of the best mapped and studied
superclusters. It is a huge conglomeration of galaxies that clearly
stands out on the sky, just south of the plane of our Galaxy on the 
northern hemisphere. The boundary of the supercluster on the northern
side is formed by the filament running southwestward from the Perseus 
cluster, a majestic chain of galaxies of enormous 
proportions. It has a length of at least $50h^{-1}\,\hbox{Mpc}$ and a
width of about $5h^{-1}\,\hbox{Mpc}$. It
even might be that the ridge extends out much further and has a total 
length of up to $140h^{-1}\,\hbox{Mpc}$, but obscuration by the
Galactic Disk prevents firm conclusions on this point. Along the ridge
we see a continuous arrangement of high density clusters and groups, 
of which the most notable ones are the Perseus cluster itself (Abell
462), Abell 347 and Abell 262. 
It happens to be one of the few large scale objects whose
structure has been studied meticulously with the help of
radiotelescopes, mainly the Arecibo telescope. Giovanelli, Haynes 
\& collaborators (see e.g. Wegner, Haynes \& Giovanelli, 1993, and
fig. 3) carried out an extensive and detailed analysis of the
structure of the whole supercluster by mapping the positions of
approximately 5000 galaxies in the region of the 
Pisces-Perseus supercluster. The determination of 21-cm redshifts for all
late-type galaxies in the region is an important and typical aspect of
their survey. Figure 2 gives an impression of the structure of the 
main Perseus-Pisces chain as obtained from this redshift 
survey (Giovanelli \& Haynes, 1996). The upper panel shows the sky 
distribution of all the survey galaxies in the Perseus-Pisces region.
The region believed to contain the PP main ridge is 
outlined. By plotting the sky position of the galaxies 
against their recession velocity $V_0$, the lower panel shows that it 
is indeed a real ridge in three-dimensional space. Notice that the Perseus
cluster is recognizable in the lower panel as the ``Finger of God''
towards the left. From figure 3 we can in fact see that it is quite 
feasible to map large scale structures with the help of for example
the SKA. Moreover, no longer would we be restricted to a 
relatively narrow redshift range and small area on the sky, so that 
mapping of structures like the Perseus-Pisces chain could be carried 
out to much higher redshifts and at virtually every location on 
a large fraction of the sky.

By now of the order 20 or so superclusters have been
identified. Claims that there might be identifiable objects of an even
larger size, in the order of several hundred Mpc, have not been 
substantiated. It is quite likely that in the local universe the
Shapley concentration, the most impressive, and monstrous, supercluster
complex that we know, represents the strongest density fluctuation on
scales of $25h^{-1}\hbox{Mpc}$ and larger. Covering an area of at
least 15 degrees in radius on the sky and
located at a distance of around $150h^{-1}\hbox{Mpc}$, some 
$100h^{-1}\hbox{Mpc}$ behind the Hydra-Centaurus supercluster, it is
the most massive concentration of clusters of galaxies in our local 
corner of the universe. The complex is dominated by a central region 
in which more than 20 rich clusters are crammed within a
significantly flattened region of radius $12.5h^{-1}\hbox{Mpc}$ around
the extremely rich cluster Shapley 8. The total mass of this crowd of 
clusters has been estimated to be in the order of $10^{17}\,M_{\odot}$ 
(see e.g. Quintana et al. 1995). Confined within such a volume this mass 
corresponds to an overdensity in excess of 2.4 times the average
density of the universe, truely exceptional for that scale. Probably 
not surprising for a region so crowded and compact, it appears to 
be dynamically very active, with clusters merging at a high rate and 
with large deviations from the Hubble flow (Raychaudhury et
al. 1991). When it turned out that the motion of our Local Group with 
respect to the universe is more or less directed towards the 
Shapley concentration, this obviously triggered speculations that its 
gravitational pull might be responsible for this peculiar
motion, and that it therefore might be the `Great Attractor' whose existence 
was inferred from the peculiar velocity field in our local
vicinity. Although this is certainly a possibility, given the 
fact that such a massive complex is located at such a nearby distance,
present estimates are that it would only account for at best 
$25\%$ of the total motion of the Local Group. 

\begin{figure}
\vspace{9.5truecm}
\caption{The Bo\"otes void as revealed by the galaxy number space 
density in a sequence of five different recession velocity intervals
in the direction of the Bo\"otes constellation on the sky. The 
lowest contour represents a density equal to 0.7 of the cosmin mean, 
each higher contour represents a factor of 2 increase in density. 
Velocity ranges ($\hbox{km/s}$): (a) 7,000-12,000; (b) 12,000-17,000; 
(c) 17,000-23,000; (d) 23,000-29,000; (e) 29,000-39,000. Frame (b) 
clearly reveals a large void in the galaxy distribution, which turns 
out to be roughly 
spherical in outline. Figure from Kirshner et al. (1987).} 
\end{figure}

\subsection{Cosmic Voids}

Perhaps one of the most intriguing discoveries emanating from
extensive redshift surveys has been the existence of large {\it voids} in the 
galaxy distribution, enormous regions, sometimes up to tens of
megaparsec in extent, wherein few or no galaxies are found. 
The Bo\"otes void in the KOSS redshift survey (Kirshner et al. 1981,
1987, see fig. 4) 
was the first void to attract the attention. It is an almost
completely empty spherical region with a diameter of around
$60h^{-1}\,\hbox{Mpc}$ and is considered to be the most typical 
example (fig. 4). Various redshift surveys covering large parts of the 
local universe have revealed that voids with sizes typically in the 
range of $\sim 20-50h^{-1}\hbox{Mpc}$ (see e.g. Vogeley, 
Geller \& Huchra 1991) are a common feature in the galaxy
distribution. This leads to the important conclusion that voids must
be an essential structural element in the universe, and recent evidence 
on the basis of a deep redshift survey (Bellanger \& de Lapparent
1995) showed that this is the case out to redshifts of at least 0.5. Also, 
redshift surveys seem to suggest that the galaxy voids are generally 
associated with surrounding enhancements in the galaxy density, and
probably these associated density excesses are stronger when 
the void is bigger. For example, the boundary on the near side of 
the Bo\"otes void is formed by the Hercules supercluster, and on the
far side by the Corona Borealis supercluster. A more detailed look at for
example the Bo\"otes void has made clear that voids are not
necessarily completely empty. They should rather be considered as 
under-populated regions of space that often do contain some galaxies. The most 
systematic effort for finding galaxies within the realm of the 
Bo\"otes void region itself is the HI survey by Szomoru et al. (1996).
In total 34 galaxies were discovered in HI, bringing the total to 58 
galaxies inside the void. Szomoru (1995) even speculated that some of 
these galaxies delineate tenuous filaments inside the void. 

That voids in the galaxy distribution also correspond to voids 
in the matter distribution is an issue that has not yet been
completely settled. In the general picture of structure formation
structure forms by the growth of initially very small density 
fluctuations under the influence of gravity (see discussion below). 
Voids in the matter distribution will have formed out of the
underdense regions in the initial matter distribution of the universe 
(see e.g. Van de Weygaert \& Van Kampen 1993). Because there 
is less matter to slow down their cosmic expansion velocity, these 
regions correspond to a lower than average gravitational attraction. The 
matter in these protovoids therefore keeps on expanding with a 
higher velocity and consequently, with respect to the expanding 
background of the universe, starts to flow out of the underdense
regions. Consequently, these protovoids will contain even less matter 
and induce an even lower gravitational attraction. This results 
obviously in an underdense region that becomes more and more empty with
time. If one were to measure the peculiar motion of matter inside and 
near the edge of these voids, one would be bound to see the relative 
expansion of the void. In other words, effectively one would observe
a `pushing' influence of the void. Indeed, studies 
of peculiar motions of galaxies in the neighbourhood of voids produced  
evidence for this gravitational influence of voids (see e.g. 
Bothun et al. 1992; Dekel \& Rees 1994). If voids indeed correspond to 
considerable perturbations in the matter distribution, they will 
represent significant disturbances in the gravitational potential.  
Using this fact, a statistical study of their sizes could yield
important constraints on the spectrum of initial perturbations 
(Blumenthal et al. 1992). Moreover, in that case the issue comes up 
whether their low density environment would 
in any way influence the process of galaxy formation. It might be 
that the formation of galaxies would be effectively suppressed 
so that they would even be more conspicuously empty in the 
galaxy distribution. Alternatively, the physical properties of 
the void galaxies may be different from the galaxies in less 
extreme and more average environments. According to some suggestions
void galaxies are more likely to be dwarf galaxies or low surface 
brightness galaxies, whose gas supply did not get efficiently
processed into stars. The implication would be that they have a
relatively high HI content, implying that further insight may come 
from 21-cm line studies. The first indications from such a survey of 
the Bo\"otes void by Szomoru et al. (1996) are that void galaxies are 
`normal'. More comprehensive and systematic studies will need the
capacity to measure HI redshifts over an extended redshift range. A
telescope like the SKA will therefore be almost essential in shedding 
more light on this issue.

\subsection{The Cosmic Foam}

Having made an inventory of the structures that we can find in the 
distribution of galaxies, we now turn to the issue of how it all fits 
together in a coherent picture. For this we return to the discussion
of redshift surveys and to the redshift map in figure 1. 
A two-dimensional ``slice'', a narrow $6^{\circ}$
band of nearly $120^{\circ}$ wide and to a depth of approximately
$15h^{-1}\hbox{Mpc}$, was taken by the CfA group as an optimal 
survey geometry for the aim of studying the morphology of 
large scale structure. The CfA2 redshift survey consists of a set  
of such slices in which the redshifts of all galaxies with an apparent
magnitude $m < 15.5$ were determined.
It was with the publication of the results of the first CfA slice, by
de Lapparent, Geller \& Huchra (1986), that we got to recognize that
the galaxies are distributed in an intriguing 
{\it foamlike} or {\it bubbly} pattern, with under-populated regions (the
{\it voids}) surrounded by {\it walls} and {\it filaments} (the 
superclusters), at whose intersections we find strong density 
enhancements in the form of {\it clusters} of galaxies (fig. 
1). When further slices were added to the first slice, the basic
picture of a `cosmic foam' got confirmed, while the existence of the 
even larger structure got revealed, the {\it Great Wall}. So far the
results for 4 slices, in total containing slightly more than 12,000
galaxies, have been published. An extension of this optical 
redshift survey on the southern sky by da Costa and collaborators (da
Costa et al. 1994), the
Southern Sky Redshift Survey extension (SSRS2) which consists of about 
3600 galaxies brighter than $m=15.5$, displays a similar pattern of 
a void-filled universe with wall-like structures surrounding empty 
voids and even of a great wall (fig. 1, da Costa 1993). 

\subsection{Theories of structure formation}

Explaining how the described intricate structural patterns in the 
galaxy distribution have originated from the almost perfectly smooth 
early universe is evidently a challenging task for theories of structure
formation. The issue of the origin of the galaxy distribution involves
two closely related problems. The first one is to explain how
structure has built up in the overall matter distribution, and the 
second one is when, how and where the galaxies have formed
in the underlying ocean of matter. An overall theory of structure
formation can only then be called complete and succesfull if it
succeeds in explaining both issues from first principles and withstand 
comparison with, among others, the observed galaxy distribution. 
Although as yet we are still far away from achieving this goal, a general 
framework for the formation of structure is gradually emerging.  

Firstly, we have to go back to the time at which we can see the 
earliest traces of structure in our universe. From the high degree of isotropy of the microwave background radiation
we can infer that the universe was highly homogeneous at the epoch of 
recombination, when electrons and protons combined into hydrogen
atoms. This occurred when the universe was a mere million years old 
and had cooled down to a temperature of $\approx 3,500\hbox{K}$, at a
redshift of $z \approx 1,300$. Since they got last scattered by free 
electrons at that epoch, the MWB photons have been travelling 
uninterruptedly until they hit our telescopes. The slight variations
in angular distribution of the measured temperature of the microwave
photon bath must therefore contain direct information on the local 
circumstances at the surface of last scattering. It was only in 1992
that such tiny anisotropies in the microwave background temperature
were actually found, when the COBE satellite detected small ripples 
in temperature with an amplitude in the order of only 
$\Delta T/T \approx 10^{-5}$ (Smoot et al. 1992). While these
anisotropies proved the existence of structures at the recombination
epoch on very large scales in the order of a gigaparsec, it also
showed how very low their amplitude is. 

The finding of COBE is a remarkable confirmation of the general
theoretical framework of `gravitational instability' for cosmic
structure formation. According
to this theory the early universe was almost perfectly smooth except for 
tiny density variations with respect to the general background 
density of the universe and related tiny velocity perturbations 
with respect to the general Hubble expansion. Because slight density 
enhancements exert a slightly stronger gravitational attraction 
on the surrounding matter, they start to accrete material from 
its surroundings as long as pressure forces are not sufficient 
to counteract this infall. In this way the overdensity becomes even 
more overdense, and their gravitational influence even stronger. The 
denser it becomes the more it will accrete, resulting in an
instability which can ultimately cause the collapse of a density 
fluctuation to a gravitationally bound object. The size and mass of 
the object is of course dependent on the scale of the fluctuation. 
For example, galaxies are thought to have formed out of 
fluctuations on a scale of $\approx 0.5h^{-1}\hbox{Mpc}$, while 
clusters of galaxies have emerged out of fluctuations on a 
larger scale of $\approx 4h^{-1}\hbox{Mpc}$. The formation of 
voids fits in the same general scheme, having grown out of 
primordial underdensities in the matter distribution. 

Providing the general framework, the gravitational instability theory 
needs lots of details to be filled in before it can be considered a 
complete theory. There is of course the issue of the amount of 
matter represented by a density fluctuation, as more 
massive fluctuations will collapse sooner. Given the amplitude of 
the fluctuations, their total mass is determined by the average 
cosmological density, paramerized by $\Omega$. The very low 
value of the amplitude of the primordial density fluctuations 
inferred from the COBE MWB measurements is a strong argument in
favour of a high overall density of the universe. Otherwise, density 
fluctuations would simply not have had sufficient time to collapse 
on all the scales that nowadays are observed to exhibit so much 
structure. Also some other observational indications support 
a high value of $\Omega$, which has the important implication 
that most likely the major share of matter in the universe does not
consist of familiar baryons and leptons but of one or 
more as yet unidentified species of `dark matter'. 

The nature and amount of dark matter is also of substantial influence
in determining the character of the initial density and fluctuation field, 
probably the most crucial issue in the structure formation saga. Rather 
than consisting of some isolated, well-defined and smooth density
peaks and dips, each of its own particular scale, the density field
can be thought of as a random superposition of fluctuations of various  
scales. It will therefore bear the character of a noise field, 
`a random field', a random superposition of waves much like the
surface of the sea at rough weather. Evidently, the waves with the 
largest amplitude will collapse first. The character of the
density field evolution will then depend on the relative
amplitudes of the different waves. One extreme case is that of
small scale waves having by far the highest amplitude. 
They will collapse into virialized objects well before a larger scale 
perturbation, in which they are possibly embedded, starts to
collapse. Consequently, we will see a hierarchical or `bottom-up'
build-up of structure, where small objects that formed first merge
into larger structures, which themselves merge to form galaxies,
cluster of galaxies, and so on. The other extreme is that of the 
case in which there are only perturbations on large scales, with no 
contributions from smaller scales. In such a `top-down' scenario 
the first emerging structures form through the collapse of those 
large scale perturbations. In the most popular versions of 
`top-down' theories these objects would correspond to superclusters. 
Subsequently, smaller objects like galaxies have to form 
through the fragmentation of these collapsed large objects into 
smaller pieces, an as yet mostly ununderstood process in which 
non-gravitational gas processes play a key role. 

The formation of anisotropic structural patterns in these random
density fields is the consequence of an additional characteristic property of 
gravitational collapse. Overdensities, on any scale and in any 
scenario, always collapse such that they become increasingly
anisotropic. At first they turn into a flattened `pancake', later 
possibly followed by contraction into an elongated filament or 
by full collapse into a virialized clump like a galaxy or a cluster. 
This tendency to collapse anisotropically is caused by the intrinsic 
primordial flattening of the overdensity as well as by the anisotropy
of the gravitational force field induced by the external matter 
distribution, i.e. by tidal forces. In the case of a pure hierarchical
scenario the amplitude of large scale overdensities will be so low
that they will not really have started their anisotropic collapse 
before the small scale overdensities have turned into high-density 
virialized clumps. Instead of appearing like a large coherent 
anisotropic structure the resulting large scale matter distribution 
will therefore more resemble a mere incoherent and shapeless density 
enhancement in the number of small clumps. On the other hand, in less 
extreme hierarchical scenarios large scale density fluctuations will
have an amplitude high enough such that by the time small scale clumps
have completely collapsed the large scale structure in which they are 
embedded will already have contracted substantially. In those cases we 
expect to see more or less coherent walls and filaments in which 
the small scale clumps stand out like beads on a string. Finally, in 
the most extreme `top-down' case we will only see the anisotropic
contraction of a large scale object like a supercluster. The resulting
pattern will be one of a network of filaments and walls without any 
internal structure. 

\begin{figure}
\vspace{18.0truecm}
\caption{The evolution of the dark matter distribution in a 
computer simulation of the gravitational growth of structure 
in the standard Cold Dark Matter scenario ($\Omega=1$ and
$H_0=50\,\hbox{km/s/Mpc}$). At six different cosmic epochs, 
corresponding to redshifts $z=2.3, 1.5, 1.0, 0.7, 0.25$ and $ 0.0$ 
(from left to right, top to bottom), the figure shows the particle 
distribution in the $5h^{-1}\hbox{Mpc}$ thick central slice 
through the simulation box of size $100h^{-1}\hbox{Mpc}$. The 
simulation follows the evolution of a system of $262,144$ 
particles in this box under the influence of their mutual 
gravity, assuming periodic boundary conditions. The N-body simulation 
was carried out with the P$^3$M gravitational N-body code of 
Bertschinger (see Bertschinger \& Gelb 1991)}. 
\end{figure}

Evidently then, the decisive factor in determining the outcome of the 
evolution of a density field, and the structure 
of the universe, is the function that specifies the relative amplitude
of the various `density' waves. This function is called the 
{\it spectrum} of density fluctuations. The present
standard view is that the very early universe produced a hierarchical 
spectrum of density fluctuations, the socalled Harrison-Zel'dovich 
spectrum. This primordial spectrum gets modified dependent on the
nature of the matter content of the density field on which these 
fluctuations are imprinted. For example, if most matter in the
universe would be dark matter in the form of massive neutrinos, the 
most typical species of hot dark matter, all fluctuations below the 
free-streaming scale of the almost relativistically moving neutrinos 
would be damped. This implies that the first perturbations that would 
collapse are on this scale, corresponding to that of present-day 
superclusters. In other words, the hot dark matter model is a top-down
scenario. However, still the most popular scenario is the Cold Dark
Matter scenario. In the CDM scenario the dark matter consists of some 
hitherto unknown species of collisionless particles having a
negligible 
velocity dispersion. Consequently, these particles will not damp any 
primordial fluctuation. The only noticeable modification of the 
primordial spectrum is a turnover at a scale of around $13/\Omega h^2 
\,\hbox{Mpc}$ caused by a suppressed growth of fluctuations 
in the epoch before matter takes over from radiation as the dominant 
component of the universe. Structure formation in the Cold Dark Matter
scenario will therefore be the result of hierarchical clustering.

Figure 5 contains an illustration of structure formation in the Cold
Dark Matter
scenario. In a box with a size of $100h^{-1}\hbox{Mpc}$ we set up a 
realization of a density and velocity fluctuation field specified by 
the CDM spectrum (for $\Omega=1.0$ and
$H_0=50\,\hbox{km/s/Mpc}$). This initial density field is then sampled by a 
large number of discrete particles. Subsequently, we let the computer 
calculate how this particle distribution would evolve due to 
their mutual gravitational interaction. In six frames (from left 
to right, top to bottom) we show the particle distribution in a 
central slice through the simulation box at successive redshifts of 
2.3, 1.5, 1.0, 0.7, 0.25 and 0.0, the present epoch. The frames
evidently show a strong growth of structure on all scales. On small
scales we see that earlier collapsed objects later merge into 
large clumps. Also noticeable is the emergence of a filamentary 
and wall-like pattern, which becomes particularly pronounced 
at later epochs due to the collapse of the corresponding 
density fluctuations and the constant flux of matter from lower
density areas. 

\begin{figure}
\vspace{14.0truecm}
\caption{The gravitational field in a N-body simulation of 
clustering in a standard Cold Dark Matter universe ($\Omega=1.0$, 
$H_0=\hbox{km/s/Mpc}$. The simulation is a $262,144$ particle 
simulation in a $100h^{-1}\hbox{Mpc}$ box with periodic boundary 
conditions. The 4 frames show, at different resolution, the particle 
distribution at redshift $z=0.0$ in 4 different $5h^{-1}\hbox{Mpc}$ 
slices. The top-left box has a size of $100h^{-1}\hbox{Mpc}$, 
the top-right one of $50h^{-1}\hbox{Mpc}$, the bottom-left one 
of $25h^{-1}\hbox{Mpc}$ and the bottom-right one of 
$15h^{-1}\hbox{Mpc}$. In addition to the particle distribution, 
each frame also displays the component of the gravity field 
in the box plane, representing it by 
a vector whose size and direction are proportional to the 
strength and direction of the gravity field at that position. 
The N-body simulation was carried out with the P$^3$M gravitational 
N-body code of Bertschinger (see Bertschinger \& Gelb 1991)}
\end{figure}

To give an idea of the dynamical background of the
formation of these features in the large scale matter distribution 
we display in figure 6 the gravitational force, wrt. the mean
expanding background of the universe, in combination with the
corresponding particle distribution. In four panels, decreasing in size from 
$100h^{-1}\hbox{Mpc}$ to $15h^{-1}\hbox{Mpc}$ as we proceed from 
the top-left to the bottom-right frame, the gravitational force at 
regularly placed locations is represented by a vector whose 
size is proportional to the strength of the gravitational acceleration
and which is directed along the force direction. In addition to the 
evident presence of strongly attracting massive clumps of matter, 
the two top panels present clear evidence for the `pushing' influence of large 
empty voids. Focussing in on the condensed matter concentrations, we 
see that while at large scales the force field looks quite regular and displays
less features than the matter distribution itself, at small scales 
it may display an interesting anisotropic infall pattern. Such an 
anisotropic force field is for example found near the agglomeration of
clumps grouped along a filament in the bottom righthand frame. 
Within the theory of gravitational instability the induced velocities 
trace very well the gravitational field up to quite a late 
stage in the development of the corresponding density fluctuation
field. This fact has spawn a large effort towards 
mapping the velocity field in the local universe, in the hope of 
being able to measure the gravity field, and thus directly the 
matter density fluctuations (see e.g. Dekel 1994 and Strauss \&
Willick 1995 for up-to-date reviews of this active and important 
cosmological field). As mentioned in section 2, the SKA could 
also yield significant advances in this scientific field through its ability
to accurately determine peculiar velocities, through 
the Tully-Fisher relation, out to much larger distances than 
hitherto possible. 

The structural pattern that is forming in the particle 
distribution in figure 5 bears a reasonably good 
resemblance to the one we observe in the galaxy distribution. 
On the other hand, careful analysis of available observational 
data suggest that available theories fail in explaining
many aspects of galaxy clustering, in particular at scales 
in the order of $100h^{-1}\hbox{Mpc}$. There are for example  
substantial observational indications for the existence of more 
power at those scales than predicted by any of the presently popular 
structure formation scenarios. Notice though that the simulation in 
figure 5 indicates that also the CDM scenario yields structures 
on a scale of $100h^{-1}\,\hbox{Mpc}$, the size of the box, in an 
advanced state of development. However, comparison of theoretical 
models with the galaxy distribution on those scales is quite hampered 
by it being near the limit of the present-day galaxy surveys. More 
definitive conclusions can therefore not be expected without exploring
the galaxy clustering out to greater depths in the universe. 
In fact, studies of the galaxy distribution at much higher redshifts  
have the additional virtue that they would possibly also yield
information on the evolution of galaxy clustering from earlier 
epochs onward. As yet the information on clustering at earlier epochs 
can at best be called scarce. 

\begin{figure}
\vspace{7.5truecm}
\caption{An example of a deep pencil beam redshift survey, showing 
the redshift distribution of galaxies out to a distance of 
of $1200h^{-1}\hbox{Mpc}$ towards the south Galactic pole (negative 
velocities) and the north Galactic pole (positive velocities). Plotted
is the number of galaxies in $10h^{-1}\hbox{Mpc}$ bins. To be precise, 
the figure is a combination of several very narrow 
pencil beam redshift surveys, comprising fields of 5 to 20
arcminutes. The black bars represent the number of galaxies in the
original survey of Broadhurst, Ellis, Koo \& Szalay (1990). The 
superposed dotted bars represent more recent extensions of and
additions to the original (1990) survey. The continuous curve at 
the background is the survey selection function, which combines the 
effects of the different geometries and apparent magnitude limits of 
composite survey beams. Figure was kindly provided by Alex Szalay.}
\end{figure}

\subsection{Exploring structures at greater cosmic depths}

While these theoretical reasons provide an important motivation for 
trying to push the galaxy redshift survey limits further out, 
the need to do so is also suggested by the present galaxy surveys themselves.  
One aspect that is quite clear from relatively shallow surveys like 
the CfA2 survey is that the largest structures that are seen are
comparable in size to the survey volume. In other words, we do not 
yet have a representative sample of structures in the universe 
and we have not yet reached the scale at which the universe 
can be considered truely homogeneous. A practical problem is that 
it becomes more and more difficult to probe the galaxy 
distribution deeper in the universe as the determination of galaxy 
redshifts demands increasing amounts of telescope time because the
galaxies become rapidly 
fainter with redshift. In order to overcome the requirements of depth,
completeness, and limited telescope time several different survey
strategies have been defined. One of the first attempts to go to 
very high redshifts restricted the survey to a very narrow angle on 
the sky. These surveys acquired the descriptive name of `pencil-beam' 
survey. A very striking redshift distribution was found in the 
pencil-beam redshift survey of Broadhurst et al. (1990). Going out as 
far as $2,000h^{-1}\hbox{Mpc}$ they found an apparent regularity in the
galaxy distribution, huge spikes separated by gaps with a size of 
$\sim 128h^{-1}\hbox{Mpc}$ (compare fig. 7, an extension of the
original Broadhurst et al. 1990 pencil beam redshift survey). It is 
rather unlikely that such large  
gaps correspond to equally large completely empty voids. The 
presence of empty voids of that size would put tight constraints 
on theoretical models of structure formation, and indeed would 
probably be in conflict with the smoothness of the microwave 
background radiation as only one such structure would be expected 
within the whole visible universe (Blumenthal et al. 1992). It is 
therefore more likely that the survey only picked up the most
conspicuous features along its line of sight while its very small 
effective beam width made it miss the smaller voids and structures in 
front and in between the big structures. Such a view is certainly 
supported by a comparison with the CfA and SSRS2 survey. It shows
that the first spikes in the pencil beam survey correspond to the 
`Great Walls' on both sides. Moreover, recent results of the equally 
deep but more completely sampled redshift survey by Bellanger \& de 
Lapparent (1995) shows that the characteristic void-filled morphology 
of the CfA2/SSRS2 survey persists out to high redshifts, with voids 
having similar diameters. Maybe the Broadhurst et al. (1990)
observation is suggestive of the existence of `supervoids', large 
underdense regions with a size of $\approx 130h^{-1}\hbox{Mpc}$ that 
are bounded by great walls and that do contain smaller voids, with sizes of 
$20-50h^{-1}\hbox{Mpc}$ in their interior. If the 
local universe indeed testifies of such a `void hierarchy' it may 
well mean that we are living in an underdense quarter of the 
universe. 

Another attempt to probe very far out is that of the sparse surveys. 
Instead of measuring all the galaxy redshifts in a specific volume of
space, only a fraction of the galaxy redshifts is measured, ideally by chosing 
the survey galaxies completely at random. They do 
not show striking structural features in as much detail as the 
dense optical surveys of CfA2 and SSRS2. However, they are 
excellent for inventarisation of structures on large scales and 
for producing maps of the galaxy density over these scales. 
An interesting example of such a survey is the QDOT survey 
(Saunders et al. 1991, QDOT is an acronym listing of the involved 
institutes, Queen Mary-Durham-Oxford-Toronto). The goal of the QDOT 
survey was to probe all structures on scales larger than 
$\sim 10h^{-1}\,\hbox{Mpc}$ within a certain distance of our Local
Group. For this purpose the survey galaxies were selected on the 
basis of the IRAS catalogue, because it provides such a well-defined 
and uniform sample of galaxies covering as much of the sky as 
possible. Important in this sense is the fact that the IRAS fluxes 
are little affected by extinction by the disk of our Milky Way. 
By its nature the IRAS catalogue already defines a sparse sample, 
as only a fraction of galaxies are represented, mainly late-type 
galaxies. From this catalogue QDOT selected 1 in every 6 galaxies 
that have a $60 {\mu\hbox{m}}$ flux brighter than 0.6 Jy, so that 
it probes out to an effective depth of $200h^{-1}\hbox{Mpc}$ over 
the largest fraction of the sky. The QDOT survey has been very useful 
in identifying large density 
enhancements and depressions within this volume, revealing the
existence of several previously unidentified voids and galaxy 
concentrations besides previously known voids and superclusters 
such as the Persues-Pisces and Coma supercluster. 

A third alternative to shortcut the effort to obtain 
information on clustering on very large scales is by concentrating 
on objects that are intrinsically more sparsely distributed than 
galaxies, that are related to certain aspects of the large scale 
matter distribution, and that can be detected out to much higher 
distances. For example, one could look at the clustering of quasars 
or other active galactic nuclei. However, their position in and 
relationship to the large scale matter distribution is still not very 
clear. Far more promising is to look at the high-density peaks of 
the galaxy distribution, i.e. at the rich clusters of galaxies. With 
their low space density and large mean separation one could compare 
their tracing of large scale structure with the way mountain peaks
trace a mountain range (Bahcall 1988). The Abell catalogue of 1682 
rich clusters, selected from the Palomar Sky Survey plates, is more or
less complete to a redshift of $z \sim 0.2$. However, most results on the large
scale distribution of clusters have been confined to a redshift 
$z < 0.1$ ($\approx 300h^{-1}\hbox{Mpc}$), based on a complete subsample of 
104 nearby rich Abell clusters for which the redshifts have been 
determined. In the meantime deeper and more objectively defined 
cluster redshift samples are being compiled. Probably 
the catalogues selected on the basis of the cluster's X-ray 
luminosity are most promising for probing very deep into the 
universe (see B\"ohringer 1995). Clusters contain a large quantity of 
hot ($10^7-10^8$ K) tenuous intracluster gas that strongly emits 
in X-rays. The total X-ray luminosity is a direct measure of the total mass 
of the cluster because of its direct relation to the depth and extent of 
the cluster potential well. Selected on the basis of the best
available X-ray imaging survey, the ROSAT All Sky Survey, the ESO Redshift
Survey of ROSAT clusters will yield upon completion the best defined and 
deepest cluster catalogue available, its $\sim 700$ southern clusters probing 
out to an effective depth of $\sim 600h^{-1}\hbox{Mpc}$ (Guzzo
1995). It might even be possible to explore scales up to 
$1,000h^{-1}\hbox{Mpc}$ ($z \sim 0.3$) on the basis of the in total 
4000 to 5000 clusters expected in the ROSAT survey.

\subsection{The new and deep galaxy redshift surveys}

While pursuing alternatives and shortcuts to complete galaxy 
redshift surveys is an efficient way to obtain some specific 
information on super large scale clustering, their information 
content cannot compete against fully and uniformly sampled galaxy 
redshift surveys. Fully sampled galaxy surveys will trace 
much better anisotropic structures like filaments and walls. 
In addition, a fully sampled galaxy redshift survey 
will yield an optimal dynamic range of scales over which the 
density fluctuation spectrum can be determined. 

A first effort towards an extension of galaxy redshift surveys 
like CfA2 and SSRS2 is the Las Campanas Redshift Survey. This 
survey includes redshifts of 26,000 galaxies with a mean redshift 
of $z \sim 0.1$, and consists of six slices covering an area 
of $\approx 700$ square degrees. Although the Las Campanas redshift 
survey has some peculiar sampling characteristics, it is well suited 
for measuring the galaxy density fluctuations at scales in the order  
of $100h^{-1}\hbox{Mpc}$. Recently, Landy et al. (1996) determined 
the power spectrum on scales between 30 and
$200h^{-1}\hbox{Mpc}$ and found that there is a strong peak at 
a scale of $\approx 100h^{-1}\hbox{Mpc}$. Furthermore, they identified
this peak to correspond to numerous walls and voids visible in the
survey. This result indicates that structures similar to the `Great
Wall' in the CfA survey and the Bo\"otes void are common features 
of the local universe, and it lends support to the claim of excess 
power on these scales inferred from the deep pencil beam survey of 
Broadhurst et al. (1990).

\begin{figure}
\vspace{10.5truecm}
\caption{A simulated slice, 6$^\circ$ by 130$^\circ$,
through the SDSS redshift survey of the north Galactic cap.  
Galaxies are plotted at the distance that would be inferred from
their redshift, so cluster velocity dispersions create ``fingers of
God'' that point towards the observer.  The slice contains 66404
galaxies, 6.6\% of the number expected over the full area of the
northern survey.  This mock catalog is drawn from a large N-body 
simulation of a low-density ($\Omega=0.4,~\Lambda=0.6$) CDM model. 
From Gott, Weinberg, Park \& Gunn 1996, also see Gunn \& Weinberg 1995.}
\end{figure}

Two truly ambitious survey projects are at the moment in an advanced 
state of preparation, both meant to probe the galaxy distribution out 
to scales of $\sim 1000h^{-1}\hbox{Mpc}$. The largest and most 
comprehensive galaxy survey is undoubtedly the Sloan Digital Sky
Survey, run by a consortium of U.S. institutions (see Gunn \& Weinberg
1995 for an extensive discussion). For this redshift project a
special, dedicated 2.5-meter telescope has been built with a
corrected field of $3^{\circ}$. Its instrumentation will include a 
large multi-CCD camera and two double fibre spectrographs of each 
320 fibres. The galaxy survey will cover a quarter of the sky in 
the northern Galactic cap, and consists of two major parts. The first 
is a large CDD
photometric survey producing images in five colours to a limiting
magnitude of $\approx 23$, yielding something like $\sim 5 \times
10^7$ galaxies. From this photometric survey all galaxies brighter 
than magnitude 18 will be selected, and of these $\sim 10^6$ 
galaxies the redshift will be determined, corresponding to a mean 
redshift of $z \approx 0.1$. In addition the survey 
will yield the spectroscopy of $\sim 10^5$ quasars, while there will 
also be a repeated imaging of galaxies in a $200\,\hbox{deg}^2$ 
strip in the southern Galactic cap. It need no saying that the 
amount of information that the Sloan survey will produce is 
truely mind-boggling. Mapping the large scale galaxy distribution will 
only be one of the applications, be it one of the more important 
ones. To give an idea of how a $6^{\circ}$ slice through the Sloan 
survey would look like, figure 8 shows a mock galaxy catalogue 
extracted from a large 54 million particle N-body simulation 
of a low-density ($\Omega=0.4,~\Lambda=0.6$) CDM model (from Gott et
al. 1996, also see Gunn \& Weinberg 1995). The slice 
contains 66404 galaxies, 6.6\% of the number expected over the full
area of the northern Sloan survey.  

A complementary survey is the Anglo-Australian 2 degree Field (2dF) 
redshift survey (see Lahav 1995), which will be carried out with the 
4 metre Anglo-Australian 
Telescope and its 400-fibre spectroscopic facility covering a
$2^{\circ}$ field of view. It will produce the redshifts for around 
$250,000$ galaxies brighter than magnitude $19.5$, selected from the 
APM survey. In total it will cover $\sim 1,700$ square degrees on the 
sky, and have a median redshift of $z \sim 0.1$. 

It is clear that a succesfull operation of the Sloan survey and 
the 2dF survey will be a giant step towards understanding structure 
formation. Obviously it will yield a large amount of information on 
density fluctuations on scales between $30h^{-1}\hbox{Mpc}$ and 
$1000h^{-1}\hbox{Mpc}$, allowing an accurate determination of the 
density fluctuation spectrum in between those scales as well as a 
better assessment of the structure of the foam-like patterns 
in the galaxy distribution and the interrelationship between 
its constituents, the galaxy walls, filaments, clusters and voids.
Moreover, hopefully this will also settle the issue at what scale 
the universe more or less reaches `homogeneity'. Besides these 
issues concerning the structure in the universe there is also 
a plethora of other, related, spin-offs. The peculiar motions of 
galaxies induced by the matter fluctuations leave their mark in 
the form of redshift distortions in the survey maps, and from 
these distortions it is possible to extract accurate estimates of 
$\Omega$, as well as to get a good idea of the biasing of the galaxy 
distribution with respect to the matter distribution. Moreover, 
just the five-colour photometric survey alone will already contain 
a wealth of information on the evolution of galaxies. 

Although these ambitious projects will provide us with these 
overwhelming amounts of information on clustering, it still concerns 
the local universe out to redshifts of $z \sim 0.1-0.2$. Going 
out to truely gigantic scales corresponding to $z \sim 1$ and 
larger would be a logical follow-up, even more so as that 
will allow for the first time exploration of the evolution of structure 
(compare e.g. fig. 5, $z \sim 0.1-0.2$ is even more recent than 
the epoch of the last but least frame). Following the specifications
for the SKA and the discussion in section 2 it appears that to 
achieve that goal the Square Kilometer Array will be THE instrument of
choice ! 

\section{Neutral hydrogen in the local universe}

As discussed in section 2, in the local universe essentially all atomic
hydrogen is associated with galaxies; there is no evidence for a major
intergalactic component of HI.
By looking at the HI content of galaxies around us one can
therefore determine the `mass-spectrum' of atomic hydrogen clouds,
and this mass spectrum can serve as a basis for determining the number
of `detectable' galaxies.
Table 1 gives the mass spectrum of HI clouds in the local Universe
as determined by Briggs (1990). For a given $S/N$
ratio, each successive line in this table requires an increase in
observing time of a factor of 10. On the basis of table 1 one would
conclude that it does not make much sense to try to detect galaxies
with $M_{HI} < 3 \times 10^9 M_{\odot}$, because the gain in the number of
galaxies is small with respect to the increase in observing
time.

\begin{table}
\begin{center}
\begin{tabular} { c c }
\hline  \\
log $M_{HI}/M_{\odot}$  &   Number per 1000 Mpc$^3$   \\
                        &   with HI mass $> M_{HI}$   \\ [0.2ex]
\hline \\
         10.5 &  \, \,    0.002              \\
         10.0 &  \, \,    0.99   \,          \\
    \,    9.5 &  \, \,    7.1    \, \,       \\
    \,    9.0 &  \,      20.6    \, \,       \\
    \,    8.5 &  \,      42.     \, \, \,    \\
    \,    8.0 &  \,      73.     \, \, \,    \\
    \,    7.5 &         115.     \, \, \,    \\
    \,    7.0 &         174.     \, \, \,    \\   [0.2ex]
\hline
\end{tabular}
\end{center}
\noindent
\caption{Cumulative number density of galaxies as a function of HI mass
(Briggs 1990).}
\end{table}
\vspace{0.5truecm}

As an illustration of the enormous potential of a SKA for studies of the
large-scale structure in the Universe, consider galaxies out to a redshift of
0.3 ({\it i.e.} six times farther then the limit in fig. 1).
Using a Hubble constant $H_{0}$ of 100 km/sec, the
distance out to this redshift is about 750 Mpc. The number of galaxies inside
this volume with a mass of atomic hydrogen larger than say $3 \times 10^{9}
M_{\odot}$ is about $ 1.3 \times 10^{7} $, which corresponds to a
surface density of 300 galaxies per degree$^2$.
A quick estimate of the observing time needed to measure line profiles
for these galaxies can be made by comparing the SKA at $z=0.3$ with the
VLA at $z=0.03$.
The performance of a SKA at $z = 0.3$ should be better or similar to that of the
VLA at $z = 0.03$, because the factor of about 80 in collecting area
improvement makes up for the increase in distance by a factor of 9.
Assuming a profile width of 120 km/sec we have $5 \times 10^{8}
M_{\odot}$ of HI per channel of 20 km/sec. Combining this with a sensitivity
of the VLA of 1.1 mJy per resolution element per hour of integration time (one
$\sigma$), a column density of HI of $5 \times 10^{8} M_{\odot}$ per
resolution element results in a 12 $\sigma$ detection in one hour.
Thus, in one hour one can obtain excellent lineprofiles with a SKA of
galaxies with $ 3 \times 10^{9} M_{\odot}$ of HI at $z=0.3$.
This translates into 7000 line profiles per square
degree per day. If one is not interested in the detailed shapes of the profiles
of the spectral lines, but only in redshifts, the observing time may be
reduced by an order of magnitude.
In other words, of order 100000 redshifts could be obtained per day
out to $z = 0.3$ with a primary beam of one square degree.
The effiency of a SKA in measuring redshifts drops rapidly with
increasing redshift, but as discussed in section 5, it will be
possible to reach redshifts around 3 or 4.

\vfill\eject
\section{Cosmological requirements for the SKA: its performance at high $z$}

The issues related to the detection of neutral hydrogen at low and high
redshifts differ fundamentally. At low $z$ essentially all galaxies will be
spatially resolved and the main questions are what column densities of HI can
be reached in a given amount of observing time, say one day, and how far one
can penetrate into the low end of the HI mass function given in table 1.
At high $z$ on the other hand the main question is whether galaxies at the upper
end of the HI mass function ($M_{HI} > 10^{9}$ to $10^{10} M_{\odot}$) can still
be detected at all, {\it i.e.} in say 100 days of observing time. This
difference comes about because, depending on the adopted cosmological model, the
observing time needed to detect a given HI mass increases at least with the
fourth power of $z$ (the total flux in an emission line decreases faster
than with $z^2$ and a decrease in the fluxdensity can only be beaten by
increasing the observing time quadratically). This indicates that for
a given HI mass there will exist a fairly sharp cut-off in $z$ beyond
which HI clouds can no longer be
observed, simply because the observing time becomes excessive.

In section 4 we have argued that it may be possible to obtain
redshifts at a rate of $10^5$ per degree$^2$ per day out to $z \simeq 0.3$.
But the real
challenge of the SKA lies at redshifts of one and beyond.
According to Braun (1995), the HI line sensitivity of a SKA at $z=1$ is close to
$1.0 \times 10^8 M_{\odot}$ ($1 \sigma$) per 24 hours observing time, per
frequency channel of 100 kHz ($\simeq 20 km/sec$) for $\Omega = 1$. This
number confirms our estimate in section 2, which translates into $0.9
\times 10^8 M_{\odot}$ ($ 1 \sigma$). For a line profile one needs say
$3 \sigma$ detections in 6 adjacent channels. The amount of hydrogen per
galaxy must therefore be at least $ 2 \times 10^9 M_{\odot}$.
Determining redshifts does not require
an entire day however: a $5 \sigma$ detection, corresponding to 2 hours, would
probably suffice. In table 2 we list the observing time needed for
measuring redshifts as a function of $z$ for a galaxy with $2 \times
10^9 M_{\odot}$ of HI per resolution element for $q_0 = 0.1$
and $q_0 = 0.5$.

\begin{table}
\begin{center}
\begin{tabular} { c @{\hspace{1.0cm}}  c @{\hspace{1.0cm}} c  }
\hline \\
$z$  & $q_0 =0.1$  &   $q_0 = 0.5$  \\ [0.2ex]
\hline \\ [0.2ex]
1.0  &  \, \,    0.17           &  \,    0.083           \\
1.5  &  \, \,    1.4  \,        &  \,    0.50  \,        \\
2.0  &  \, \,    6.6  \,        &  \,    1.8   \, \,     \\
2.5  &  \,      22.   \, \,     &  \,    5.0   \, \,     \\
3.0  &  \,      62.   \, \,     &       11.    \, \, \,  \\
3.5  &         148.   \, \,     &       23.    \, \, \,  \\
4.0  &         316.   \, \,     &       41.    \, \, \,  \\  [0.2ex]
\hline
\end{tabular}
\end{center}
\caption{Observing time needed for redshift determinations for galaxies
with $2 \times 10^9 M_{\odot}$ of HI (in days).}
\end{table}
\vspace{0.5truecm}

Let us now try to estimate the number of galaxies inside a spherical
shell at a given redshift that might be detected by a SKA. This estimate
will be based on the mass function of HI in galaxies given in table 1.
It is of course unlikely that this mass function will be applicable at
high redshift. At least three processes affect this mass function.
The conversion of gas into stars in the course of time will
increase the average HI content of galaxies with increasing look-back
time. At present the HI content of most of the brighter galaxies does
not exceed 5 to 10 \% of the stellar mass. Thus, at high $z$, depending
on the star formation history, the typical HI mass per galaxies could be much
higher than it is now. On the other hand, accretion of HI onto galaxies,
the merging of small galaxies into bigger ones, and the higher
level of ionising radiation at high $z$, act in the opposite
sense. Little is known about these processes. Accretion and merging
could be quite
important because the faint end of the galaxy luminosity function is
known to evolve strongly.

In table 3 we estimate the number of detectable galaxies per square
degree. Following Braun (1995), this estimate is based on a receiver
with 1280 channels. Using channels of 100 kHz ($\simeq 20$ km/sec) at
the emission frequency of 1421 MHz, results in the redshift interval
given in the second column of table 3. In practice the situation is more
favourable than our estimate indicates because the primary beam of the
telescope will open up at larger wavelengths. For a primary beam of one
degree$^2$ at 21 cm, the number of galaxies per beam is a factor of
$(1+z)^2$ larger than that given in the table.

\begin{table}
\begin{center}
\begin{tabular} { c  @{\hspace{1.0cm}}  c c c }
\hline \\
$z$  & $z$ interval    &   $N_{gal}$ $q_0=0.1$ & $N_{gal}$ $q_0 = 0.5$  \\
     &                 &   $\times 1000$ & $\times 1000$      \\ [0.2ex]
\hline \\[0.2ex]
1.0  &  0.92 -- 1.10   & \, \,    5.8      &  \,    3.8       \\
1.5  &  1.39 -- 1.62   & \,      14.  \,   &  \,    7.8       \\
2.0  &  1.87 -- 2.14   & \,      27.  \,   &       13.   \,   \\
2.5  &  2.35 -- 2.67   & \,      46.  \,   &       19.   \,   \\
3.0  &  2.83 -- 3.19   & \,      69.  \,   &       26.   \,   \\
3.5  &  3.31 -- 3.72   & \,      99.  \,   &       33.   \,   \\
4.0  &  3.79 -- 4.24   &        134.  \,   &       42.   \,   \\ [0.2ex]
\hline
\end{tabular}
\end{center}
\caption{Estimated number of galaxies with $M_{HI} > 3 \times 10^9
M_{\odot}$ per square degree.}
\end{table}
\vspace{0.5truecm}

At large redshifts it will in general not be possible to resolve the detected
objects spatially. However, it is extremely important that the surroundings of
galaxies can be mapped at least coarsely, with a linear resolution of say 5 kpc,
in order to allow a study of the building up of galaxies from smaller units.
For a large range in redshifts 5 kpc corresponds to an angular resolution of
about one arcsec, and the resolution elements of a SKA should certainly be no
smaller than this. The spectral resolution needed for cosmological studies is
modest. With channels of about 20 km/sec wide one can obtain rough information
on the shape of the line profiles and redshifts can be determined with an
accuracy of a few km/sec.

In summary, studies of HI at high redshifts demand the following:

\begin{itemize}
\item Primary beam: as large as possible. Beams smaller than about one square
degree will degrade the potential of the SKA for studies beyond $z = 1$
substantially.
\item Resolution elements: about one arcsec.
\item Spectral resolution: channels of about 20 km/sec (at emission).
\end{itemize}

\begin{acknowledgements}
We wish to thank Hans B\"ohringer, Frank Briggs and Luiz da Costa 
for encouraging and useful discussions. We are very grateful to 
Luiz da Costa, Hans B\"ohringer, Martha Haynes, Alex Szalay and David Weinberg 
for providing respectively figure 1, figure 2, figure 3, figure 7 and figure 
8. In addition, we are grateful to J.R. Gott for the permission to use 
figure 8. Also we wish to acknowledge Doris Neumann for critically 
reading the text and for many helpful comments and suggestions, 
and Ed Bertschinger for providing the P$^3$M N-body code.
RvdW is supported by a fellowship of the Royal Netherlands Academy of 
Arts and Sciences. He also acknowledges the hospitality of the 
Max-Planck-Institut f\"ur Astrophysik in Garching, Germany, where part of this
contribution was written. 

The Digitized Sky Survey, used to obtain figure 2, was produced at the Space
Telescope Science Institute under US Government grant NAG W-2166, and 
is based on photographic data of the National Geographic Society -- Palomar
Observatory Sky Survey, funded by a grant from the National
Geographic Society to the California Institute of Technology.
\end{acknowledgements}

\end{document}